# Investigating Academic Major Differences in perception of Computer Self-efficacy and Intention toward E-learning Adoption in China

Nattaporn Thongsri[a], Liang Shen[b], Yukun Bao[a*]

[a] Center for Modern Information Management,
School of Management, Huazhong University of Science and Technology, Wuhan, P.R.China, 430074; [b] Center for Big Data Analytics, Jiangxi University of Engineering, Xinyu, Jiangxi Province, P.R.China, 338000

**Abstract**

Recognizing the underlying relationship between e-learning practice and the institutional environments hosted in, the Chinese educational practice on branching high school students into science, technology, engineering, and mathematics (STEM) and non-STEM academic major groups before being admitted into universities or colleges is examined. By extending the well-established Technology Acceptance Model (TAM) with computer self-efficacy, this study aims to examine the difference in perceptions and behaviours on e-learning adoption from the STEM and non-STEM students. The results revealed that STEM's score of computer self-efficacy, perceived ease of use and behavioural intention to use e-learning are all greater than non-STEM's.

**Keywords:** Academic major difference, STEM, e-learning, TAM, Self-efficacy, China

* Corresponding author: Tel: +86-27-87558579; fax: +86-27-87556437.
Email: yukunbao@ hust.edu.cn or y.bao@ieee.org

**Introduction**

The Chinese higher education system has expanded rapidly since 1980, with more than 2000 higher education institutions (Liu & Wang, 2015). E-learning is a learning process that delivers content and interaction between learners and teachers through electronic media and is a term that is synonymous with blended learning or online learning (Kibelloh & Bao, 2014a). Learning through e-learning can remove the need of a classroom, which provides convenience, flexibility, and destroys the boundary of traditional learning. Changing traditional learning styles to e-learning is a current issue for Chinese education, as China has not yet fully embraced e-learning (Duan, He, Feng, Li, & Fu, 2010). Thus, it is interesting to see why some learners choose e-learning, while others do not, and this is an important motivation for this study. Specifically, the issue addressed in this study arises from the common practice of separating high school students into science, technology, engineering, and mathematics (STEM) and non-STEM academic major groups before being admitted into universities or colleges, this division of academic major is a growing concern due to the possible impact it may have on the decision making process to use e-learning.

Understanding the factors that influence the use of e-learning has been the focus of numerous e-learning studies in recent years (Bao, Xiong, Hu, & Kibelloh, 2013; Kibelloh & Bao, 2014b; Thongsri, Shen, Bao, & Alharbi, 2018). A factor that has not been considered, in technology acceptance literature, is the role of the academic major (i.e. STEM or non-STEM) for computer self-efficacy and subsequent e-learning adoption. This suggests, that technology users akin to e-learners assumedly possess the same skill set, knowledge and attitude towards e-learning regardless of academic major background, which is misleading. In addition, Veenstra, Dey, and Herrin (2008) found that students in different majors had different perceptions regarding computer self-efficacy. Chiou and Liang (2012) found that maths and science related barriers and supports correlated significantly with self-efficacy, coping efficacy, and outcome expectations. Furthermore, Bandura (1986) theorized that individual perceptions of self-efficacy are derived from the four dimensions of information, which are: mastery experiences, vicarious learning, social persuasion, and physiological states. Each of these four sources was reported to correlate with mathematics self-efficacy (Lent, Lopez, & Bieschke, 1991). Though STEM majors, may or may not necessarily be directly related to computers; the overall STEM curriculum provides knowledge and skills (i.e. problem solving and analytical skills) derived from mathematics, which give self-confidence and lead to effective performance in related domains. However, the issue of academic major differences is especially relevant in China, given that students are streamed into STEM and non-STEM majors from as early as high school. This early academic major filtering is arguably a critical part in the formulation of the Chinese STEM 'leaky pipeline' due to the number of STEM graduates is very few (Clark Blickenstaff, 2005). This phenomenon is undoubtedly a source of confidence and self-efficacy concern toward technology use, which certainly merits further research.

**Theoretical development and hypothesis**

*Research model*

The Technology Acceptance Model (TAM) consists of two main variables: perceived usefulness and perceived ease of use (Davis, 1989). Despite the credit being given to the TAM model for its ability to explain and predict behavioral intention and user behavior regarding technology, there are some primary limitations relevant to the implementation of the current study. In the case of e-learning, these variables that derive from the theory cannot be used very effectively in the study of student motivation, and therefore, they should be studied along with additional variables as well (Ong & Lai, 2006). This study argues explicitly that a TAM extension is needed to illuminate deep-rooted academic major difference issues related to effects on-learning.

In understanding people's attitude and behavior in using online applications and services, self-efficacy has been investigated in various contexts, such as internet self-efficacy (Eastin & LaRose, 2000), or web-specific self-efficacy (Hsu & Chiu, 2004). This study uses computer self-efficacy as suggested by Compeau and Higgins (1995) to reflect and operationalize different hardware and software configurations.

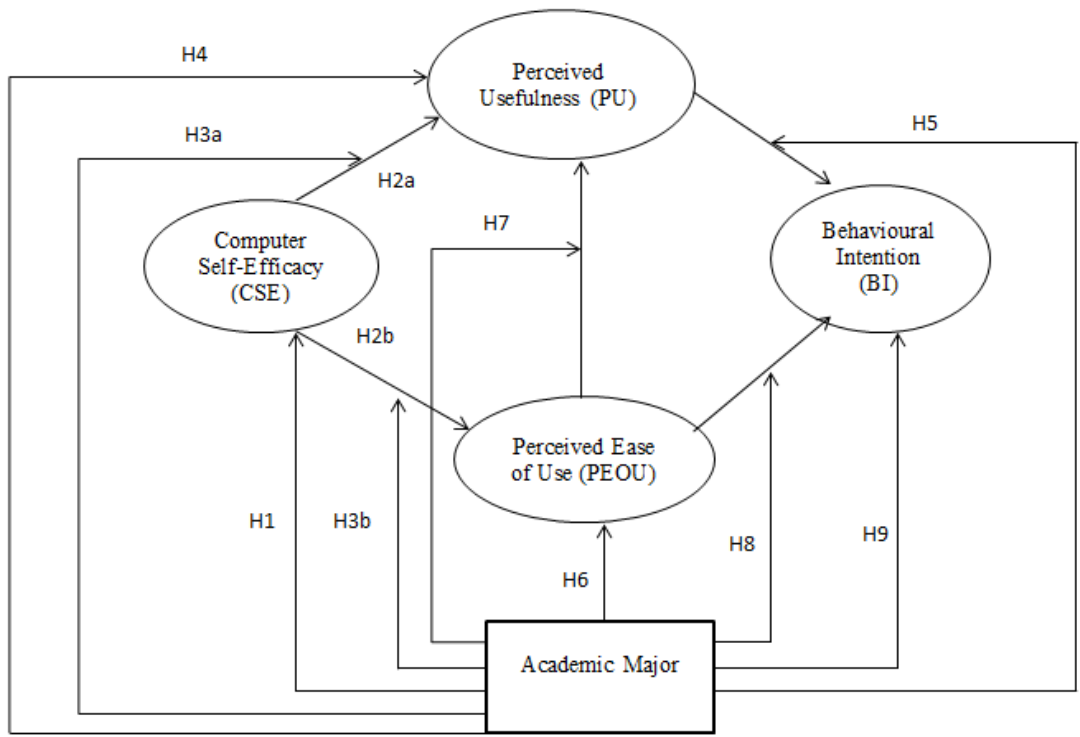

**Figure 1:** The proposed research TAM model, with extended variable computer self-efficacy.

**Computer self-efficacy and academic major**

Bandura (1986) states that the ability of a person to evaluate is based on a perception of past experience or is the ability of an individual to assess his or her ability to achieve the desired goal. This study associated prior experience with general skills gained from STEM majors, regardless of whether they are directly linked with computers. Self-efficacy beliefs are important to motivate users to realize and perceive performance in their previous tasks, and the ability to perform subsequent behaviours in the future (Vogel & Human-Vogel, 2016). In accordance with the STEM curriculum is the development of technology-based activities to promote the problem-solving process. This will result in students in STEM education familiar with using a computer fairly well(Lee, 2002). Moreover, China has set policies, promote online learning and establish the National Distance Education Collaboration Group of Chinese Higher Education Institutes, the objective is to share resources and encourage online learning. Therefore, students who are familiar with the technology tend to choose the STEM fields when they enter the university because they believe that technology-based activities will give them the skills, familiarity, and understanding of the nature of STEM (Cherney, 2008; Lee, 2002).Thus, the following hypothesis was proposed:

H1: STEM students' rating on CSE is higher than non-STEM students.

*Computer self-efficacy and TAM variables*

The majority of previous studies attempted to test the statistical relationships between computer self-efficacy and perceived ease of use in order to investigate the causal relationship between both variables (Ong & Lai, 2006; Venkatesh & Davis, 1996). Therefore, previous findings emphasize the relationship between computer self-efficacy and perceived ease of use of e-learning. If the user has a positive perception of computer self-efficacy, this will also affect the perception that e-learning is easy to use and that users do not have to exert a large amount of effort when using the system. Furthermore, evidence was empirically examined, and it was found that there appears to be a relationship between self-efficacy and perceived usefulness that is if the user has a positive perception of computer self-efficacy, this will also affect the perception that e-learning is beneficial in helping learners achieve academic goals (Compeau &

Higgins, 1995). Taken together the aforementioned differences between STEM and non-STEM students, therefore, the current study's hypotheses are the following:

H2a: CSE will have a positive effect on PU.

H2b: CSE will have a positive effect on PEOU.

H3a: CSE will influence PU more strongly for non-STEM students than STEM students.

H3b: CSE will influence PEOU more strongly for non-STEM students than STEM students.

*Perceived usefulness*

Many previous studies have confirmed the relationship of the path to perceived usefulness to behavioural intention (Davis, 1989; Kibelloh & Bao, 2014a). STEM students out of interest and experience may find technology such as e-learning more useful than non-STEM. Thus, the following hypotheses were proposed:

H4. STEM students' rating on PU is higher than non-STEM students.

H5. PU influences BI more strongly for STEM students than for non-STEM students.

*Perceived ease of use*

Although the perceived ease of use is only as an indirect determinant of behaviour intention, many empirical studies have found strong relationships between perceived usefulness and behavioural intention (Venkatesh & Davis, 2000). Whilst perceived ease of use is hypothesised to have a direct impact on perceived usefulness, the reverse is not true. This is because perceived usefulness is a variable that is used to study the benefits of a system, which results in the user being able to accomplish tasks or systems that help to make the job completed more efficiently, while the perceived ease of use is the variable used to study the user-friendliness of systems, which directly affects job performance (Davis, 1989). The above discussion concludes that PU and PEOU together have a significant effect on BI. In addition, PEOU will positively affect both PU and BI by testing the causal relationship of PEOU on PU repeatedly to confirm the correlation between the two variables (Davis, 1989; Venkatesh, 1999). Due to the varying backgrounds and course curriculum, it is expected students of STEM majors to have a competitive advantage over students' non-STEM majors in terms of technology skills and rich computer-related experiences. Thus, the current study's hypotheses are the following:

H6. STEM students' rating on PEOU is higher than non-STEM students.

H7. PEOU influences PU more strongly for non-STEM students than for STEM students.

H8. PEOU influences BI more strongly for non-STEM students than for STEM students.

*Behavioural intention to use*

In this study, the dependent variable employed was the intention to use e-learning, as it is closely related to actual use (Hu, Clark, & Ma, 2003). Actual use is the behaviour while the intention to use is the attitude. In the study of behavioural use, there are complex relationships through causal relationships (Seddon, 1997; Talukder, Chiong, Bao, & Hayat Malik, 2018). For this reason, the study through the intention to use e-learning variable has several advantages due to actual use possibly having several causes and delicate factors, which may not result in a comprehensive study (Teo, 2011). It is expected students of STEM majors to have a stronger willing of usage of computer-related technology over students of non-STEM majors. Thus, the following hypotheses were proposed:

H9. STEM students' rating of BI is higher than non-STEM students.

**Methodology**

The STEM and non-STEM definitions used in the current research were based on a review of the standardized Chinese universities curriculum set by the Ministry of Education of the People's Republic of China. The questionnaire consisted of two parts, a section for responses to the research constructs and another to measure demographic variables: gender, age, previous education background, education level and current academic major. Content validity was ensured by all of the four items; computer self-efficacy, perceived usefulness, perceived ease of use, behavioural intention being adopted from previous validated tools (Compeau & Higgins, 1995; Ong & Lai, 2006; Venkatesh & Davis, 1996). Four hundred and fifty questionnaires were distributed to the students; there were 432 completed questionnaires (i.e. all items were answered) equalling 97.10%.

**Table 1.** Demographic profiles and descriptive statistics of respondents.

| Characteristics | Frequency | Percent |
|---|---|---|
| *Gender* | | |
| Male | 228 | 52.78 |
| Female | 204 | 47.22 |
| *Age* | | |
| less than 18 | 2 | 0.46 |
| 18-20 | 200 | 46.30 |
| 21-23 | 147 | 34.03 |
| 23-25 | 83 | 19.21 |
| *Majors* | | |
| *STEM* | | |
| Computer science | 40 | 9.26 |
| Telecommunications | 34 | 7.87 |
| Mechanics | 39 | 9.03 |
| Natural science | 31 | 7.18 |
| Information systems | 44 | 10.19 |
| Electronics | 32 | 7.41 |
| Civil | 17 | 3.94 |
| *Number of STEM students* | *237* | *54.86* |
| *Non-STEM* | | |
| Management business administration | 38 | 8.80 |
| Marketing | 32 | 7.41 |
| Human resource | 22 | 5.09 |
| Public administration | 27 | 6.25 |
| Finance | 31 | 7.18 |
| Education | 21 | 4.86 |
| Economics | 24 | 5.56 |
| *Number of non-STEM students* | *195* | *45.14* |

**Data analysis and results**

**Assessment of measurement model**

Structural Equation Modelling (SEM) based on the two-step approach introduced by Anderson and Gerbing (1988) was used. First, confirmatory factor analysis (CFA) was used to develop the measurement model, and second, we analysed the structural model to test the hypotheses. AMOS 17.0 was used for data analysis by the method of confirmatory factor analysis to test the measurement model.

To assess the model's overall goodness of fit six common model-fit measures were used: the ratio of $\chi^2$ to degrees of freedom ($df$), the goodness –of-fit index (GFI), adjusted goodness-of-fit index (AGFI), normalized fit index (NFI), comparative fit index (CFI) and root mean square residual (RMSR) ($\chi2/df$ =1.921, GFI = 0.869, AGFI = 0.809, NFI = 0.921, CFI = 0.917, RMSR = 0.036 ). This study has shown a good fit between the model and data. All the model-fit indices exceeded their respective common acceptable levels suggested in Gefen, Straub, and Boudreau (2000). This study has shown a good fit between the model and data. All the model-fit indices exceeded their respective common acceptable levels suggested in Gefen et al. (2000).

**Table 2.** Fit indices for measurement and structural models

| Goodness-of-fit measures | Recommended value | Measurement model | Structural model |
|---|---|---|---|
| $\chi2$ /degree of freedom | $\leq 3$ | 1.921 | 1.974 |
| Goodness-of-fit index (GFI) | $\geq 0.8$ | 0.869 | 0.880 |
| Adjusted goodness-of-fit index (AGFI) | $\geq 0.8$ | 0.809 | 0.830 |
| Normed fit index (NFI) | $\geq 0.9$ | 0.921 | 0.970 |
| Comparative fit index (CFI) | $\geq 0.9$ | 0.917 | 0.980 |
| Root mean square residual (RMSR) | $\leq 0.05$ | 0.036 | 0.041 |

Reliability and validity were used to test the measurement model by considering the values of composite reliability and average variance extracted. Reliability was measured by two indices: Cronbach's $\alpha$ and composite reliability (CR). All factors in this study's measurement model had composite reliability greater than 0.90. The average extracted variances were all greater than the recommended 0.50 level (Hair & Hair, 1992), meaning, for all variance observed in the items, more than one-half was accounted for by hypothesized factors.

Convergent validity was based on factor loadings and squared multiple correlations with the method of confirmatory factor analysis. Following the recommendations of Hair and Hair (1992), a factor loading and squared multiple correlations should greater than 0.5 was considered to be very significant. The results show that the values of reliability and convergent validity in this research are good.

Discriminant validity was determined by the average variance extracted. Compared to its relationship with other levels, the square root of the AVE should be superior (Fornell & Larcker, 1981). In Table 3, the bold items shows the squares root of the AVEs for the constructs as accepted.

**Table 3.** Reliability, average variance extracted and discriminant validity

| Factor | CR | 1 | 2 | 3 | 4 |
|---|---|---|---|---|---|
| 1. Computer Self-efficacy | 0.960 | **0.793** | | | |
| 3. Peceived usefulness | 0.906 | 0.410 | **0.742** | | |
| 4. Perceived ease of use | 0.963 | 0.336 | 0.395 | **0.707** | |
| 5. Behavioural intention to use | 0.970 | 0.422 | 0.430 | 0.255 | **0.746** |

Diagonal elements are the average variance extracted. Off-diagonal elements are the shared variance. CR is composite reliability.

**Table 4.** Factor loadings and squared multiple correlations of items

|  | Factor loadings | Squared multiple correlations |
|---|---|---|
| Computer self-efficacy | | |
| CSE1 | 0.621 | 0.786 |
| CSE2 | 0.758 | 0.874 |
| CSE3 | 0.634 | 0.801 |
| CSE4 | 0.656 | 0.808 |
| Perceived usefulness | | |
| PU1 | 0.842 | 0.709 |
| PU2 | 0.719 | 0.617 |
| PU3 | 0.682 | 0.665 |
| PU4 | 0.691 | 0.778 |
| Perceived ease of use | | |
| PEOU1 | 0.676 | 0.741 |
| PEOU2 | 0.655 | 0.807 |
| PEOU3 | 0.859 | 0.738 |
| PEOU4 | 0.735 | 0.840 |
| Behavioural Intention | | |
| BI1 | 0.909 | 0.825 |
| BI2 | 0.816 | 0.866 |

*Structural model*

Similar six common model-fit indices were used to examine the structural model ($\chi^2/df$ =1.974, GFI = 0.880, AGFI = 0.830, NFI = 0.970, CFI = 0.980, RMSR = 0.041). The results suggested that the measurement model showed a reasonably good fit with the collected data. Consequently, this research could proceed with investigating the determinants and mediating effects of academic major difference in the intention to use e-learning.

*Hypothesis testing*

The Statistical Package for the Social Sciences (SPSS), a statistical analysis technique based on ANOVAs, aims to study the impact of major differences on e-learning intentions through CSE, PU, and PEOU. The mean scores, standard deviation, together with significant *F* ratios, are shown in Table 5.

The structural model defines the casual path relationship of variables in the proposed model. In order to test the hypotheses, the significance of difference was calculated using the procedure described in Cohen, Cohen, West, and Aiken (2013). Properties of the causal paths, including standardized path coefficients and the significance of difference in the hypothesized model, are presented in Table 6 and Table 7 summarizes the testing results of hypotheses. Within the STEM students perceived usefulness show highest "direct" and "total" effect on behaviour intention to use. However, the result founded computer self-efficacy had influence on intention to use through perceived usefulness and perceived ease of use for non-STEM students higher than STEM students. The direct, indirect and total effect of CSE, perceived usefulness and perceived ease of use on behavioural intention to use for academic majors difference are summarized in Table 8.

**Table 5.** Descriptive statistics and ANOVAs testing results

|      | STEM (n=237) | | Non-STEM (n=195) | | Significance of difference between STEM and Non-STEM $F$ ratios |
|------|------|------|------|------|------|
|      | **Mean** | **SD** | **Mean** | **SD** | |
| **CSE**  | 6.16 | 1.03 | 4.16 | 1.26 | 12.17*** |
| **PU**   | 5.54 | 0.77 | 5.29 | 0.92 | ns |
| **PEOU** | 5.67 | 0.86 | 4.21 | 0.91 | 13.02*** |
| **BI**   | 5.83 | 1.05 | 5.08 | 1.16 | 5.74* |

ns, not significant; * $P<0.05$; ***, $P<0.001$.

**Table 6.** Academic major differences in relationships of CSE-PU, CSE-PEOU, PU-BI, PEOU-PU, and PEOU-BI

|  | Entire Sample β | STEM (n=237) β | NON-STEM (n=195) β | Difference between STEM and NON-STEM |
|---|---|---|---|---|
| CSE-PU | 0.057* | 0.025$^{ns}$ | 0.410$^{ns}$ | * |
| CSE_PEOU | 0.336* | 0.196* | 0.640* | * |
| PU-BI | 0.717* | 0.815$^{ns}$ | 0.595* | * |
| PEOU-PU | 0.276** | 0.151$^{ns}$ | 0.366* | ** |
| PEOU-BI | 0.043* | 0.094* | 0.055* | ns |

ns, not significant;* P < 0.05;** P < 0.01.

**Table 7.** Summary of testing results

|  | Relationship | Hypothesis | Result |
|---|---|---|---|
| **Main effect** | | | |
| H2a | CSE-PU | Positive | Supported |
| H2b | CSE-PEOU | Positive | Supported |
| | | | |
| **Academic major difference** | | | |
| H3a | CSE-PU | Non-STEM > STEM | Supported |
| H3b | CSE-PEOU | Non-STEM > STEM | Supported |
| H5 | PU-BI | STEM > Non-STEM | supported |
| H7 | PEOU-PU | Non-STEM > STEM | Supported |
| H8 | PEOU-BI | Non-STEM > STEM | Not supported |
| | | | |
| **Perception** | | | |
| H1 | CSE | STEM> Non-STEM | Supported |
| H4 | PU | STEM > Non-STEM | Not supported |
| H6 | PEOU | STEM > Non-STEM | Supported |
| H9 | BI | STEM > Non-STEM | Supported |

**Table 8.** Academic major differences between the direct and indirect effects of CSE, PU, PEOU and BI

|  |  | Entire sample | | | STEM (n=237) | | | NON-STEM (n=195) | | |
|---|---|---|---|---|---|---|---|---|---|---|
|  |  | PU | PEOU | BI | PU | PEOU | BI | PU | PEOU | BI |
| Direct effects | CSE | 0.057 | 0.336 |  | 0.025 | 0.196 |  | 0.410 | 0.640 |  |
|  | PU |  |  | 0.717 |  |  | 0.815 |  |  | 0.595 |
|  | PEOU | 0.276 |  | 0.043 | 0.151 |  | 0.094 | 0.366 |  | 0.055 |
| Indirect effects | CSE | 0.093 |  | 0.122 | 0.030 |  | 0.062 | 0.235 |  | 0.418 |
|  | PU |  |  |  |  |  |  |  |  |  |
|  | PEOU |  |  | 0.198 |  |  | 0.123 |  |  | 0.218 |
| Total effects | CSE | 0.150 | 0.336 | 0.122 | 0.055 | 0.196 | 0.062 | 0.645 | 0.640 | 0.418 |
|  | PU |  |  | 0.717 |  |  | 0.815 |  |  | 0.595 |
|  | PEOU | 0.276 |  | 0.241 | 0.151 |  | 0.217 | 0.366 |  | 0.273 |

**Discussion & implications**

*Computer self-efficacy and e-learning acceptance*

This study attempts to add to the insights of previous research by studying computer self-efficacy of students through two main variables: perceived usefulness and perceived ease of use from the Technology Acceptance Model to study the factors that affect the acceptance of e-learning by students in STEM and non-STEM programs. The findings of this study are discussed below, and present valuable implications for the study of e-learning, especially for learners who are in different majors. The results of this study are consistent with Ong and Lai (2006), who found that computer self-efficacy, social perceived ease of use, and perceived usefulness have a statistically significant effect on e-learning acceptance.

It was also found that computer self-efficacy has the highest influence on acceptability. This finding supports the belief that if learners are competent or effective in using computers to help with the learning process, they are more likely to accept online learning such as e-learning than learners whose ability to operate a computer is low. The results also show that if a student has a positive perception of computer self-efficacy, it will also result in them realizing the benefits of learning through e-learning. Therefore, this research underscores the link between perceived usefulness as the main variable associated with self-efficacy in the use of the system. Therefore, in the development and design of an e-learning system, due to it being an educational system, system designers should focus on the development of functions and content, as well as valuable content for learners. In order to respond and promote the benefits of using the system, system designers should consider the accuracy and availability of information in order to maximize the user's experience (Wang, Wu, & Wang, 2009). The results of this study support previous research which found that, if users have a positive perception of computer self-efficacy, it will make them aware that the use of the system is easy to operate, and that they can solve the problems immediately when experiencing problems in using the system. Therefore, this research confirms the relationship between computer self-efficacy and PEOU (Thong, Hong, & Tam, 2002; Venkatesh & Davis, 2000). In addition, this research confirms the relationship between perceived ease of use and system usage; if the learner recognizes that the e-learning system is easy to use, they can operate independently of the effort and this will encourage students to use the system. The results of this study are consistent with the research of Selim (2003). Empirical studies indicate that the development of e-learning should focus on ease of use in order to attract learners to use the system and should also emphasize the development of Graphic User Interfaces (GUI) that are user-friendly, application navigations, frequently asked questions (FAQs), and user manuals. E-learning system designers can develop interfaces that allow users to customize systems according to their learning preferences. However, computer self-efficacy in each context is different depending on the characteristic of usage; user perception may change when used continuously. More research should investigate the perception of students who have used e-learning practically, to understand the influence of computers self-efficacy on the effectiveness of learning. The results of the study may provide a different perspective from the current one.

*Academic major difference*

Based on the above results, it is apparent that the role of computer self-efficacy can be used to investigate and predict the computer self-efficacy to study the adoption behaviour of educational technology. In particular, this study highlights the differences in computer self-efficacy in the acceptance of e-learning by the learners. This study found that STEM students have computer self-efficacy that is higher than that of non-STEM students. The results indicate that the different disciplines affect the students' ability to use the computer. In addition, the results show that the direct effect of computer self-efficacy on perceived ease of use is higher than the direct effect of computer self-efficacy on perceived usefulness in both STEM and non-STEM learners.

However, when considering the total effects, the results suggest that the computer self- efficacy on perceived ease of use is greater than that of computer self-efficacy towards perceived usefulness only for STEM students. Thus, it is concluded that STEM Students are more likely to use e-learning in their studies, while non-STEM students are less likely to use e-learning. Based on these findings, educational policymakers can modify non-STEM curriculums to reflect courses and trainings that can not only improve the learning experiences of students, but also ensure that future students are capable of effortlessly integrating new technologies into their learning practices. Moreover, knowledge and attitudes especially of teachers in information technology can in turn influence the development of positive student attitudes, interest and confidence toward the field (Hains-Wesson & Young, 2017). Therefore, teacher training to support students' technology use and self-efficacy increase would also be beneficial.

This study also provided that non-STEM's perception of usefulness was the most significant direct effect than STEM's in determining intention to take up e-learning in the present study. The benefits of the system are also important for online learning. Developers must prioritize content to cover the learning objectives in order to fulfil the needs of the learners. While in practice, e-learning is more field-specific, the practical aspect of STEM subjects requiring lab experiments or writing formulas such as chemistry and mathematics calculus, may cause students to find it difficult to comprehend the usefulness of e-learning. However, justifying and validating this explanation needs further investigations in future studies. More research is needed on effectiveness of designing, developing and packaging STEM content to appeal to the perceived usefulness of STEM students. E-learning educators should take advantage of the value-adding characteristics of e-learning for STEM majors in promoting perceived usefulness. For example, it seems to be a good strategy to emphasize that e-learning can help individuals simulate real-life projects online and increase their competitive advantage in learning.

**Conclusion**

The contributions of this study to e-learning acceptance research are as follows: first, this research complements the understanding of previous research by introducing the computer self-efficacy variable to be studied in conjunction with the Technology Acceptance Model in order to study the e-learning acceptance of students in China by adding academic major differences as a moderator. It was also found that computer self-efficacy affects the intended use of e-learning at a statistically significant level. Specifically, empirical studies emphasize that differences in subject matter affect intention. We found that STEM scores are higher than non-STEM students. However, regarding non-STEM students, this study found that computer self-efficacy has a statistically significant impact on the perceived usefulness of e-learning that is higher than that of STEM students. Overall, the findings reveal that non-STEM students are at risk of missing out on the benefits of e-learning. More research in this area is warranted, with follow up interviews or focus groups to gain an in-depth understanding of the adoption of e-learning in China.

**Acknowledgement**

This study was supported by the Natural Science Foundation of China (71810107003).

.